\documentclass[runningheads]{llncs}
\usepackage{graphicx}
\usepackage{times}
\usepackage{amsmath}
\usepackage{amssymb}
\usepackage{bm}
\usepackage{color}
\usepackage[ruled,vlined]{algorithm2e}
\usepackage{multirow}
\usepackage{fancyhdr}
\fancypagestyle{specialfooter}{%
  \fancyhf{}
  
  \fancyhead[R]{\scriptsize{Accepted in CVIP 2021 Conference}}
  \fancyfoot[L]{\scriptsize{$^2$Work done while at IIIT Sri City}}
}

\begin{document}
\title{AC-CovidNet: Attention Guided Contrastive CNN for Recognition of Covid-19 in Chest X-Ray Images}
\titlerunning{AC-CovidNet}

\author{Anirudh Ambati$^1$, Shiv Ram Dubey$^2$}
\authorrunning{Ambati et al.}

\institute{$^1$Indian Institute of Information Technology, Sri City, Chittoor\\
$^2$Indian Institute of Information Technology, Allahabad\\
anirudh.a17@iiits.in, srdubey@iiita.ac.in}

\maketitle

\thispagestyle{specialfooter}

%%%%%%%%% ABSTRACT
\begin{abstract}
Covid-19 global pandemic continues to devastate health care systems across the world. 
% In many countries, the $2^{nd}$ wave is very severe. Economical and rapid testing, as well as diagnosis, is urgently needed to control the pandemic. 
At present, the Covid-19 testing is costly and time-consuming. Chest X-Ray (CXR) testing can be a fast, scalable, and non-invasive method. The existing methods suffer due to the limited CXR samples available from Covid-19. 
Thus, inspired by the limitations of the open-source work in this field, we propose attention guided contrastive CNN architecture (AC-CovidNet) for Covid-19 detection in CXR images. The proposed method learns the robust and discriminative features with the help of contrastive loss. Moreover, the proposed method gives more importance to the infected regions as guided by the attention mechanism. We compute the sensitivity of the proposed method over the publicly available Covid-19 dataset. It is observed that the proposed AC-CovidNet exhibits very promising performance as compared to the existing methods even with limited training data. It can tackle the bottleneck of CXR Covid-19 datasets being faced by the researchers. The code used in this paper is released publicly at \url{https://github.com/shivram1987/AC-CovidNet/}.
\end{abstract}

%%%%%%%%% BODY TEXT
\section{Introduction}
Coronavirus disease 2019 (Covid-19) has emerged very fast as an emergent health risk disease affecting the whole world. 
% It is caused by Novel Coronavirus that leads to Pneumonia-like symptoms, including cough, fever, and/or difficulty in breathing. 
It has been observed that this infection spreads through the surfaces which might be infected from the infected person. 
% Thus, once the spread rate increases, it becomes uncontrollable due to a chain reaction like infection spread. 
The spread of Covid-19 is categorized into different stages. The stage1 and stage2 refer to the small scale spread, whereas stage3 and beyond refer to the large scale spread due to chain reaction.
% As per the World Health Organisation (WHO), Covid-19 was first observed on $31^{st}$ December 2019 as the reported case from Wuhan city of China. The outbreak was declared a Public Health Emergency of International Concern on $30^{th}$ January 2020. Countries like the USA, India, Brazil, Italy, and Spain are the worst affected ones with a huge number of infected cases. 
Covid-19 pandemic is being witnessed as the toughest time of the century during April-May 2021 due to its $2^{nd}$ wave which has already entered into stage3/stage4 (i.e., community spread) of Covid-19 infection spread. 
% As per the Covid-19 data released by WHO\footnote{https://covid19.who.int/} on $23^{rd}$ April 2021, the total number of confirmed cases has reached 143,445,675 with 3,051,736 deaths globally as of $22^{nd}$ April 2021.
Thus, the pandemic has led to a huge burden on the healthcare systems across the world.
% As suggested by various professionals and healthcare agencies, the most efficient way to tackle this problem is to adapt mass testing, contact tracing, and isolation. 
Testing for Covid-19 is the most important part of the process and must be scaled as much as possible. CXR based testing can one of the fastest way using existing infrastructure and can be scaled very quickly and cost effectively. 
% It has been observed so far that this virus attacks the human respiratory system including the lungs which is responsible for taking in oxygen and expelling out carbon dioxide. Thus, leading to breathing difficulty in the patients\footnote{https://www.who.int/health-topics/coronavirus}. 
The radiograph image of the lungs can be captured using different imaging tools such as CT-Scan and X-Ray. Getting the CT-Scan is again a costly and time-consuming process. Moreover, only major hospitals have CT scanners. However, capturing an X-Ray is a very affordable as well as an efficient process.
The covid-19 disease is caused by severe acute respiratory syndrome coronavirus-2 (SARS-COV-2) and the infected patients show distinct visual features in the Chest X-Ray (CXR) images. Hence, artificial intelligence based automated techniques can be utilized to detect the infection in CXR images. Such testing methods can be fast, scalable, economical, and affordable.
 
Researchers have tried to explore the artificial intelligence based deep learning techniques for Covid-19 detection, such as COVID-Net \cite{Wang2020}, CovidAID \cite{covidaid}, COVID-CAPS \cite{afshar2020covidcaps}, CovXNet \cite{covxnet}, DarkCovidNet \cite{darkcovidnet} and Convolutional Neural Network (CNN) Ensemble \cite{cnnensemble}. The lack of sufficient data to train and test the models is the main problem in the development of the deep learning based models for Covid-19 detection in CXR images. Hence, it is an urgent requirement to develop a deep learning model that would learn distinctive features from the limited data. In order to learn the discriminative and localized features, we propose an attention guided contrastive CNN (AC-CovidNet) for Covid-19 recognition from CXR images. Following are the commitments of this work:
\begin{itemize}
    \item We propose a novel AC-CovidNet deep learning framework for Covid-19 recognition in CXR images.
    \item The use of the attention module enforces the learning of localized visual features corresponding to Covid-19 symptoms.
    \item The contrastive loss increases the discriminative ability and robustness of the model by learning the similarity between Covid-19 infected samples and dissimilarity between Covid-19 positive and negative samples.
    % \item The proposed AC-CovidNet model is able to learn the task specific important features from the limited training data.
    \item The impact of the proposed method is analyzed for different amount of training data w.r.t. the recent state-of-the-art models.
\end{itemize}

The remaining paper is organized as follows: Section \ref{relatedworks} summarizes the related works; Section \ref{proposedmethod} illustrates the proposed AC-CovidNet model; Section \ref{experimentalsetup} details the experimental settings; Section \ref{experimentalresults} presents the results and analysis; and finally Section \ref{conclusion} summarizes the findings with concluding remarks.

\section{Related Works}
\label{relatedworks}
It has been observed in the primary research conducted by Wang and Wong (2020) \cite{Wang2020} that chest radiograph images can be used for the Covid-19 detection. It opened up the new urgent and demanding area of the possible usage of Artificial Intelligence for early, efficient and large-scale detection of viruses among people.
% Some companies around the world working in AI area have started working over AI based solutions for the Covid-19 detection using CT scans and X-Ray radiograph images\footnote{https://spectrum.ieee.org/the-human-os/biomedical/imaging/hospitals-deploy-ai-tools-detect-covid19-chest-scans}. They have started harnessing the power of deep learning to learn the infection pattern. 
% An update over the essentials of Radiologists on Covid-19 is published by Kanne et al. \cite{kanne2020essentials}. 
Ng et al. released the imaging profile of the Covid-19 infection with radiologic findings \cite{ng2020imaging}.
Li et al. discovered the spectrum of CT findings and temporal progression of Covid-19 disease which reveals that this problem can be solved using imaging AI based tools \cite{li2020coronavirus}.
Bai et al. performed a performance study of radiologists which can differentiate the Covid-19 from viral pneumonia on chest CT \cite{bai2020performance}. 
% Residual network based deep learning model \cite{farooq2020covid}, truncated inception network \cite{das2020truncated}, and deep learning based anomaly detection \cite{zhang2020covid} are also used for the screening of Covid-19 from radiographs. 
% The uncertainty and interpretability are estimated by Ghosal and Tucker \cite{ghoshal2020estimating} in the deep learning models for Covid-19 detection.
% Deep CNN model is also employed by Narin et al. \cite{narin2020automatic} to perform the automatic detection of Covid-19 from X-Ray images. 

Deep learning is also utilized for Covid-19 detection from Chest X-Ray (CXR) images \cite{narin2020automatic}.
In one of the first attempts, CXR radiograph images are used for Covid-19 detection using a Deep Learning based convolutional neural network (CNN) model COVID-Net \cite{Wang2020}.
A projection expansion projection extension module is used heavily in COVID-Net which is experimented on various configurations of the model. Authors used a human-machine collaborative design strategy to create COVID-Net architecture where human driven prototyping and machine based exploration is combined. A COVIDx dataset is also accumulated from various sources and being updated with new data. The dataset and models are publically released for further research\footnote{https://github.com/lindawangg/COVID-Net/}. Using COVID-Net model, 96\% sensitivity is observed in \cite{Wang2020} on a test set of 100 CXR images. 
% \subsubsection{CovidAID}
A CovidAID model is proposed in \cite{covidaid} which is a pretrained CheXNet \cite{rajpurkar2017chexnet} - a 121 layer DenseNet \cite{huang2018densely} followed by a fully connected layer. Using the CovidAID model, 100\% sensitivity is observed on a test set having 30 Covid-19 images. 
% \subsubsection{COVID-CAPS}
A capsule networks based deep learning model (COVID-CAPS) is investigated in \cite{afshar2020covidcaps}. Authors in \cite{afshar2020covidcaps} aim to prevent the loss of spatial information which is observed in CNN based methods. Using COVID-CAPS model, a sensitivity of 90\% is reported on 100 test CXR images.
% \subsubsection{CovXNet}
CovXNet model \cite{covxnet} is proposed to use transferable multi-receptive feature optimisation. Basically, 4 different configurations of a network are utilized in CovXNet for training and prediction. Using CovXNet model, 91\% sensitivity is reported on 100 test CXR images.
% \subsubsection{DarkCovidNet}
The DarkNet architecture based DarkCovidNet model is introduced in \cite{darkcovidnet} with You Only Look Once (YOLO) real time object detection system for Covid-19 detection. Using the DarkCovidNet model, a classification accuracy of 98.08\% for binary classes and 87.02\% for multi-class cases are observed.
% \subsubsection{Covid-19 detection using Ensemble Learning CNNs}
CNN ensemble of DenseNet201, Resnet50-v2 and Inception-v3 is utilized for Covid-19 recognition in \cite{cnnensemble}. Samples from only 2 classes (i.e., covid and non-covid) are used to train the CNN ensemble model. Using CNN ensemble model, a classification accuracy of 91.62\% is reported. Researchers from \cite{he2020sample} tried to develop an open source framework of algorithms to detect covid19 using CT scan images. Also, researchers at \cite{javaheri2020covidctnet} developed Covid-CT using selftrans approach. We have tested these algorithms on our CXR dataset configurations.

From the above presented works, it is convincing that AI powered deep learning methods can play a vital role for the Covid-19 screening. The CT scans and CXR radiographs are used majorly for the imaging based techniques. Less attention has been given to CXR images so far due to the not so great generalization performance caused by the limited availability of data \cite{ahmed2021discovery}, \cite{degrave2021ai}.
Given the need to conduct the mass screening at affordable cost, the further fast research over CXR images of lungs is very much needed using the limited data. Thus, in this paper, we utilize the capability of attention mechanism and contrastive learning to tackle the learning with limited data for Covid-19 recognition from CXR images.

\section{Proposed AC-CovidNet CNN Model}
\label{proposedmethod}
In this section, first we provide a brief of deep learning, attention mechanism and contrastive learning. Then we present the proposed AC-CovidNet architecture.
% along with objective function.

\subsection{Background}
Deep learning has shown a great impact from last decade to solve many challenging problems \cite{lecun2015deep}. Deep learning models consist of the deep neural networks which learn the important features from the data automatically. The training of the deep models is generally performed using stochastic gradient descent optimization \cite{adam}, \cite{diffgrad}. Convolutional neural network (CNN) based models have been used to deal with the image data, such as image classification \cite{alexnet}, face recognition \cite{hmloss}, image retrieval \cite{dubey2020decade}, hyperspectral image analysis \cite{hybridsn}, and biomedical image analysis \cite{lbpdad}. 

Attention mechanism in deep learning facilitates to learn the localized features which is more important in the context of the problem of Covid-19 recognition from CXR images \cite{bahdanau2015neural}. It is also discovered that the attention based model can outperform the plain neural network models \cite{vaswani2017attention}. The attention mechanism has been also utilized for different applications such as facial micro-expression recognition \cite{gajjala2020meranet}, breast tumor segmentation \cite{vakanski2020attention}, and face recognition \cite{rao2017attention}. Thus, motivated from the success of attention mechanisms, we utilize it in the proposed model for Covid-19 recognition.

Contrastive learning is the recent trend to learn the similarity and dissimilarity between the similar and dissimilar samples in the abstract feature space for visual representations \cite{chen2020simple}, \cite{supcon}. Generally, contrastive learning is dependent upon the feature similarity between positive pairs and negative pairs \cite{tian2020makes}. The contrastive learning has shown very promising performance for different problems, such as face generation \cite{deng2020disentangled}, image-to-image translation \cite{park2020contrastive}, medical visual representations \cite{zhang2020contrastive}, and video representation \cite{qian2020spatiotemporal}. Thus, motivated from the discriminative and robust feature representation by contrastive learning, we utilize it in the proposed method.

\begin{figure}[!t]
    \centering
    \includegraphics[width=8cm]{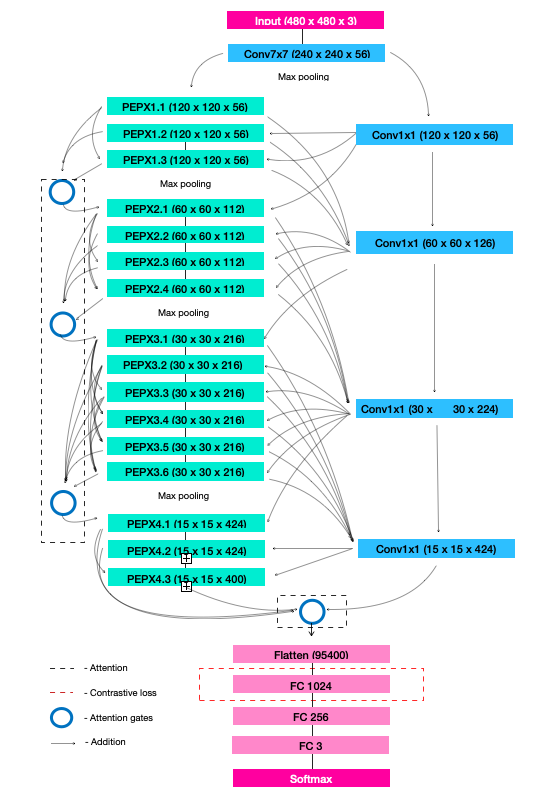}
    \includegraphics[trim={0 40 0 40},clip, width=4.1cm]{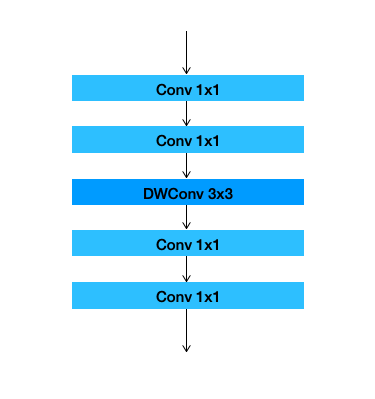}
    \caption{Left: Architecture of the proposed AC-CovidNet. Right: Projection Expansion Projection Extension (PEPX) Module. Here, DWConv representsthe depth wise convolution operation.}
    \label{fig:proposed_model}
\end{figure}

\subsection{Proposed Model}
In this paper, we propose an attention guided contrastive CNN for Covid-19 recognition, named as AC-CovidNet. The architecture of the proposed AC-CovidNet model is illustrated in Fig. \ref{fig:proposed_model}. The proposed model is based on the popular COVID-Net model \cite{Wang2020}. It heavily uses light weight residual projection expansion projection extension (PEPX) mechanism. The PEPX component is shown in Fig. \ref{fig:proposed_model} (right side). This architecture also uses selective long range connectivity in model which improves the representational capacity. It also facilitates the training of the model easier. However, the extensive use of these long range connections may bring a lot of redundant low level features. In order to resolve this issue, the proposed model uses an attention mechanism. Attention helps the model to prioritize the regions of important w.r.t. the problem being solved. Attention is also useful to suppress the activations from the redundant features from the initial layers and helps to focus on the important features that are required to solve the given problem. We use the attention gates in the proposed architecture as suggested in \cite{oktay2018attention}, at various layers in the COVID-Net architecture where many long range connections are used. This improves the sensitivity as the model attends better to the important visual features of infected regions in CXR images due to Covid-19.
As the difference between Covid-19 and Pneumonia features is very subtle, we propose to use the supervised contrastive loss. The contrastive loss facilitates the network to increase the distance between the learnt representation of the classes as much as possible.

\subsubsection{Architecture}
The proposed AC-CovidNet architecture is an extension of COVID-Net by utilizing the attention gates where multiple long range connections converge. The details of the architecture of the proposed model is illustrated Fig. \ref{fig:proposed_model}.
We use PEPX layers and attention gates heavily as a part of our architecture. We also use the contrastive loss function while training.
The model is first trained using supervised contrastive learning before final fine tuning using supervised learning.

% \begin{figure}[!t]
%     \centering
%     \includegraphics[trim={0 40 0 40},clip, width=4.5cm]{PEPX.png}
%     \caption{Projection Expansion Projection Extension (PEPX) Module. Here, DWConv represents the depth wise convolution operation.}
%     \label{fig:pepx}
% \end{figure}

\subsubsection{PEPX Layer}
The architecture of a Projection-Expansion-Projection-Extension (PEPX) is shown in Fig. \ref{fig:proposed_model} (right side). The idea of this module is to project features into a lower dimension using the first two conv1x1 layers, then expand those features using a depthwise convolution layer (DWConv3x3) and project into lower dimension again using two conv1x1 layers. Thus, the PEPX layer leads to the efficient model by reducing the number of parameters and operations.

\subsubsection{Attention Gate}
We use attention gates in the proposed AC-CovidNet model at various layers as depicted in Fig. \ref{fig:proposed_model}. The block diagram of the attention module is shown in Fig. \ref{proposedmethod}. Features from multiple layers are passed through conv1x1 and are added together. Then, the aggregated features are passed through Relu activation function followed by conv1x1 and then sigmoid activation function. The feature map output of the sigmoid layer is then passed through a resampler. The output of the resampler is added to the features from the nearest input layer to the attention module in order to produce the output of the attention gate.

\begin{figure*}[!t]
\centering
\includegraphics[trim={0 58 0 48},clip,width=12cm]{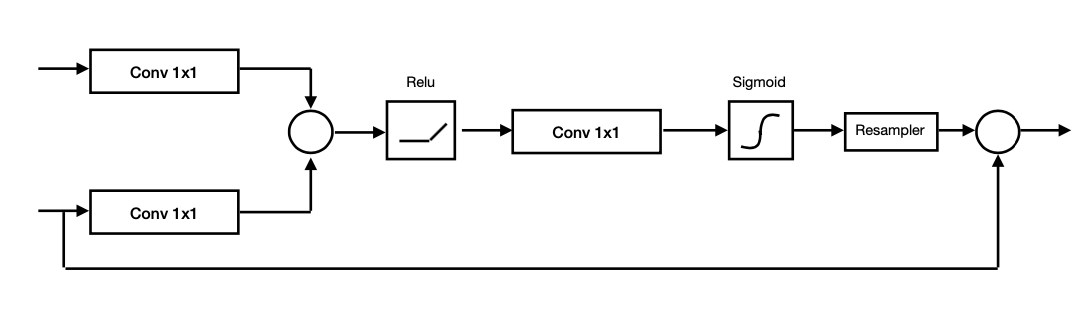}
   \caption{The block diagram of an attention gate.}
\label{fig:short}
\end{figure*}

\subsubsection{Network Details}
The following are the main components of the proposed AC-CovidNet model. a) Encoder network, $E(.)$: It is the AC-CovidNet model as discussed above after removing the last three layers which form the classifier network. The encoder network maps the input image into a vector of size 1024. b) Projection network, $P(.)$: This network projects the output of the encoder network into a vector of size 128. This is a multilayer perceptron with input size of 1024, one hidden layer of length 512, and an output vector of size 128. This network is used while training the encoder and later discarded. c) Classifier network, $C(.)$: This is the last three layers of AC-CovidNet that take a 1024 length output of the encoder as the input and produces an output of size $c = 3$ (corresponding to the classes, Normal, Pneumonia, and Covid-19). 

First we train the encoder network $E(.)$ with supervised contrastive loss (SupConLoss) using the projection network $P(.)$. Then we discard the projection network and add the classifier network $C(.)$ to the encoder network. Then we freeze the weights of the encoder model and train the classifier with categorical cross entropy loss. The training of the encoder of the proposed model is summarized in Algorithm \ref{algo_encoder}.

\begin{algorithm}[!t]
\label{algo_encoder}
\SetAlgoLined
$E$: Encoder Network, $P$: Projection Network\\
$C$: Classifier Network\\
$SupCon$: Supervised Contrastive Loss\\
$CrossEntropy$: Cross Entropy Loss\\
\While{$epochs--$}{
\tcp{Stage 1}
\For{batch $X$ in $Data$}{
    Initialize $Z$ as null\;
    \For{$x_i$ in $X$}{
        $h_i = E(x_i)$\;
        $z_i = P(h_i)$\;
        $Z.append(z_i)$\;
    }
    $\mathcal{L}_{SC} = SupCon(Z_i)$\;
    Update $E$ and $P$ to minimise $\mathcal{L}_{SC}$\;
}
}
Discard $P(.)$ and freeze weights of $E(.)$\;
Final model, $M(\boldsymbol{x}) = C(E(\boldsymbol{x}))$\;
\While{$epochs--$}{
\tcp{Stage 2}
\For{batch $X$ in $Data$}{
$\hat{Y} = M(X)$\;
$\mathcal{L}_{CE} = CrossEntropy(Y, \hat{Y})$\;
Update $M$ to minimise $\mathcal{L}_{CE}$\;
}
}
 \caption{Training AC-CovidNet}
\end{algorithm}

\subsection{Objective Function}
We use the supervised contrastive learning method \cite{supcon} to train the encoder network for feature extraction. We train the classifier network using the cross entropy loss function after freezing the encoder network.

\subsubsection{Supervised Contrastive loss}
Contrastive loss is most commonly used in unsupervised and self-supervised learning. In order to adapt this method to supervised learning and take advantage of the labels available, the supervised contrastive learning has been investigated in \cite{supcon}. This loss is used to train the encoder network of the proposed AC-CovidNet. 

Consider $\boldsymbol{z}_{\ell}=\operatorname{P}\left(\operatorname{E}\left({\boldsymbol{x}}_{\ell}\right)\right)$ where ${\boldsymbol{x}}$ is the input and $A(i)$ is a set of all indices ($I$) except $i$. Then, the supervised contrastive loss is given as,
\[\mathcal{L}_{SC}=\sum_{i \in I} \frac{-1}{|P(i)|} \sum_{p \in P(i)} \log \frac{\exp \left(\boldsymbol{z}_{i} \cdot \boldsymbol{z}_{p} / \tau\right)}{\sum_{a \in A(i)} \exp \left(\boldsymbol{z}_{i} \cdot \boldsymbol{z}_{a} / \tau\right)}\]
where, $\tau \in \mathcal{R}^{+}$ is scalar temperature parameter, $P(i) \equiv\left\{p \in A(i): \tilde{\boldsymbol{y}}_{p}=\tilde{\boldsymbol{y}}_{i}\right\}$ is the set of indices of all the positive samples other than $i$, and $|P(i)|$ is its cardinality. Note that $\tilde{\boldsymbol{y}}$ represents the class label.

\subsubsection{Cross Entropy Loss}
Cross entropy loss is used to train the classifier which takes input from the feature extractor. It produces the output probability corresponding to three class from the softmax activation function. The cross entropy loss is given as,
\[\mathcal{L}_{CE} = -\frac{1}{N}\sum_{i=1}^{\mathrm{N}}\sum_{j=1}^{\mathrm{3}} y^j_i \cdot \mathrm{log}\; {\hat{y}}^j_i\]
% \[\mathcal{L}_{CE} = -\frac{1}{N}\sum_{j=1}^{\mathrm{N}} \log \frac{e^{s^c_i}}{\sum_{j=1}^{k}{e^{s^j_i}}} \]
where ${\hat{y}}^j_i$ and $y^j_i$ are the output probability and ground truth value, respectively, for $j^{th}$ class corresponding to $i^{th}$ sample and $N$ represents the number of samples in a batch.
% where $s^c_i$ is the class score generated by the network w.r.t. the correct class for the $i^{th}$ sample, $s^j_i$ is the class score generated by the network w.r.t. the $j^{th}$ class for the $i^{th}$ sample and $N$ is the total number of samples in a batch. 

\begin{table*}
\caption{Datasets configurations used for training and testing w.r.t. the number of samples in Covid-19, Pneumonia and Normal categories.}
\centering
\resizebox{12cm}{!}{
\begin{tabular}{|c|c|c|c|c|c|c|c|}
\hline
\multirow{2}{*}{S. No} & \multirow{2}{*}{Configuration} & \multicolumn{2}{c|}{Covid-19} & \multicolumn{2}{c|}{Pneumonia} & \multicolumn{2}{c|}{Normal} \\ \cline{3-8} 
 &  & \#Training Images & \#Test Images & \#Training Images & \#Test Images & \#Training Images & \#Test Images \\ \hline
I & COVIDx-v1 & 517 & 100 & 7966 & 100 & 5475 & 100 \\ \hline
II & COVIDx-v2 & 467 & 150 & 7916 & 150 & 5425 & 150 \\ \hline
III & COVIDx-v3 & 417 & 200 & 7866 & 200 & 5375 & 200 \\ \hline
\end{tabular}}
\label{table:dataset}
\end{table*}

\begin{table*}[!t]
\caption{The results comparison in terms of the sensitivity.}
\resizebox{12cm}{!}{
\begin{tabular}{|l|l|l|l|l|l|l|l|l|l|}
\hline
\multicolumn{1}{|c|}{Model} & \multicolumn{3}{c|}{COVIDx-v1} & \multicolumn{3}{c|}{COVIDx-v2} & \multicolumn{3}{c|}{COVIDx-v3} \\ \hline
\multicolumn{1}{|c|}{} & \multicolumn{1}{c|}{Covid} & \multicolumn{1}{c|}{Pneumonia} & \multicolumn{1}{c|}{Normal} & \multicolumn{1}{c|}{Covid} & \multicolumn{1}{c|}{Pneumonia} & \multicolumn{1}{c|}{Normal} & \multicolumn{1}{c|}{Covid} & \multicolumn{1}{c|}{Pneumonia} & \multicolumn{1}{c|}{Normal} \\ \hline
COVID-Net \cite{Wang2020} & 96 & 89 & 95 & 94.66 & 87.33 & 95.33 & 94 & 88.5 & 94 \\ \hline
CovidAID \cite{covidaid} & 93 & 95 & 73 & 91 & 92.66 & 70.66 & 88 & 92.5 & 74 \\ \hline
Ensemble Learning CNNs \cite{cnnensemble} & 89 & 91 & 87 & 87.33 & 92 & 86 & 83 & 91 & 89 \\ \hline
DarkCovidNet \cite{alexnet} & 86 & 88 & 92 & 84.66 & 84 & 89.33 & 84.5 & 89.5 & 90.5 \\ \hline
Covid-Caps \cite{afshar2020covidcaps} & 90 & 87 & 77 & 88 & 86 & 74 & 86.5 & 84.5 & 76 \\ \hline
CovidCTNet \cite{javaheri2020covidctnet} & 83 & 76 & 89 & 79 & 74.33 & 87.66 & 79 & 74.5 & 88 \\ \hline
Covid-CT \cite{he2020sample} & 87 & 82 & 86 & 85.66 & 79.33 & 85 & 85 & 77 & 85.5 \\ \hline

CovXnet \cite{covxnet} & 91 & 71 & 86 & 89.33 & 69.33 & 86.66 & 88 & 70 & 84 \\ \hline
CovidNet + attention & 95 & 86 & 95 & 95.33 & 82.66 & 95.33 & 94 & 87 & 94.5 \\ \hline
CovidNet + contrastive loss & 96 & 89 & 95 & 96 & 87.33 & 94 & 95 & 90 & 95 \\ \hline
\textbf{AC-CovidNet} & \textbf{96} & \textbf{88} & \textbf{95} & \textbf{96.66} & \textbf{87.33} & \textbf{95.33} & \textbf{96.5} & \textbf{87.5} & \textbf{95} \\ \hline
\end{tabular}}
\label{table:results}
\end{table*}

\section{Experimental Setup}
\label{experimentalsetup}

\subsection{Dataset Used}
We use COVIDx dataset \cite{Wang2020} in this work to experiment with the proposed AC-CovidNet model. It is the largest open-source dataset for Covid-19 Chest X-Ray (CXR) images currently available. This dataset is a combination of several other open datasets, including Covid-19 Image Data Collection, Covid-19 Chest X-ray Dataset Initiative, ActualMed Covid-19 Chest X-ray Dataset Initiative, RSNA Pneumonia Detection Challenge dataset, and Covid-19 radiography database. 
This dataset contains three classes of CXR images, i.e., Covid-19, Pneumonia and Normal. It has a total of 14,258 images from 14,042 patients. It has 7,966 Normal images, 5,475 Pneumonia images and 517 Covid-19 images in the training set and 100 images from each category in the test set. In our experiments we create three different sets of data by varying the distributions of samples in training and test sets. We name the original COVIDx dataset as COVIDx-v1 by following the above said distribution. COVIDx-v2 is the COVIDx dataset with 150 test images from each category and 7,916 Normal images, 5,425 Pneumonia images and 467 Covid-19 images in the training set. Similarly, COVIDx-v3 has 200 test images from each category and 7,866 Normal images, 5,375 Pneumonia images and 417 Covid-19 images in the training set. These configurations are summarized in Table \ref{table:dataset} w.r.t. the number of samples in Covid-19, Pneumonia and Normal categories.

\subsection{Training Settings}
The proposed AC-COVIDNet model is pretrained on imagenet as suggested by \cite{Wang2020}. Then the entire model is trained in two stages. In the first stage, the encoder network (i.e., feature extractor) is trained using contrastive loss function as suggested in \cite{supcon} for feature extraction, so that the distance between the learnt features is optimized. In the second stage, the feature extractor is frozen and trained by adding a classifier on the top with cross entropy loss function. 
We train the proposed model as well as the state-of-the-art models on all three variations of COVIDx dataset, i.e., with 100, 150 and 200 test images of different categories. The models are trained using Adam optimiser \cite{adam}. The learning rate is set as 1.7e-4. The batch size of 64 is used. We use Relu activation function in every layer of the network and softmax in the last layer. Max-pooling is used after every batch of PEPX layers. The three versions of the COVIDx dataset are used for training and the sensitivity of the classes are compared. Covid-19 sensitivity is the percentage of instances with Covid-19 that are correctly identified. The model is trained using the computational resources provided by Google Colab. Keras deep learning library is used with tensorflow as a backend. 

\section{Experimental Results and Analysis}
\label{experimentalresults}

We test the proposed AC-CovidNet model on all three configurations of the COVIDx dataset and calculate the sensitivity for Covid-19 and other classes. In order to demonstrate the superiority of the proposed method, we also compute the results using state-of-the-art deep learning based Covid-19 recognition models, such as CovXNet \cite{covxnet}, COVID-CAPS \cite{afshar2020covidcaps}, CNN Ensemble \cite{cnnensemble}, DarkCovidNet \cite{darkcovidnet}, COVID-Net \cite{Wang2020}, and CovidAID \cite{covidaid}.
The results in terms of the sensitivity for the Covid-19 class are reported in Table \ref{table:results}.
On configuration I (i.e., COVIDx-v1 dataset with 100 test images), configuration II (i.e., COVIDx-v2 dataset with 150 test images), and configuration III (i.e., COVIDx-v3 dataset with 200 test images), the observed Covid-19 sensitivity using the proposed AC-CovidNet model is 96\%, 96.66\%, and 96.5\%, respectively. 
It can be observed in Table \ref{table:results} that the proposed model outperforms the remaining models over all three settings of the COVIDx dataset.

% We have also tested the existing models of COVID-Net \cite{Wang2020} and CovidAID \cite{covidaid} against all three configurations of the dataset. The corresponding results along with those from AC-CovidNet has been tabulated in Table 2.

% \subsection{Analysis of the results}

Note that the proposed model is able to achieve better results than the other compared models because the proposed model learns the Covid-19 specific features using the attention module and increases the separation between different classes in feature space using the contrastive loss. In the configuration I of the dataset (with 100 test images), the performance of the proposed model is better than CovXNet, COVID-CAPS, CNN Ensemble, DarkCovidNet and CovidAID models and same as COVID-Net. Thus, in order to demonstrate the advantage of the proposed model, we compare the results with less number of training samples and more test samples. Basically, it depicts the generalization capability of the proposed model. On an expectation, a bigger test set can reflect the better generalization of the deep learning models. Thus, we experiment with configuration II having 467 training samples and 150 test samples of Covid-19 category. It can be seen in Table \ref{table:results} that the proposed model is able to retain the similar performance by correctly classifying 145 Covid-19 images out of 150 in the dataset. However, the performance of other models dropped significantly. We also test the performance by further reducing the number of training samples and increasing the number of test samples in configuration III with 417 training images and 200 test images from Covid-19 category. It can be noticed that the performance of the proposed AC-CovidNet model is still similar. However, the other models drastically fail to generalize in case of limited training set. Thus, it clearly indicates that the proposed AC-CovidNet model is able to capture the robust and discriminative features pertaining to the Covid-19 infection and generalize well even with the limited training data. It also shows the positive impact of attention modules and contrastive learning for the Covid-19 recognition from CXR images.

The impact of the attention and contrastive mechanism is also investigated by considering the only attention and only contrastive mechanism with base network CovidNet (i.e., CovidNet + attention and CovidNet + contrastive loss, respectively). As shown in the results using these methods in Table \ref{experimentalresults}, the performance of AC-CovidNet, which uses both attention and contrastive mechanisms, is improved than the models which use only attention and only contrastive mechanisms for Covid-19 recognition. We also report the results for other two classes in Table \ref{experimentalresults}, i.e., Pneumonia and Normal. It is observed that the performance of the proposed model is comparable to the state-of-the-art for Pneumonia and Normal classes.

\section{Conclusion}
\label{conclusion}
In this paper, an AC-CovidNet model is proposed for Covid-19 recognition from chest X-Ray images. The proposed model utilizes the attention module in order to learn the task specific features by better attending the infected regions in the images. The proposed model also utilizes contrastive learning in order to achieve the better separation in the feature space by increasing the discriminative ability and increasing robustness. The results are computed over three different configurations of Covid-19 dataset with varying number of training and test samples. The results are also compared with six recent state-of-the-art deep learning models. It is noticed that the proposed AC-CovidNet model outperforms the existing models in terms of the sensitivity for the Covid-19 category. Moreover, it is also observed that the performance of the proposed model is consistent with a limited training set. Whereas, the existing methods fail to do so. It shows the better generalization capability of the proposed method. The future work includes the utilization of recent development in deep learning to solve the Covid-19 recognition problem from chest X-Ray images with better performance.

\bibliographystyle{splncs04}
\bibliography{refs}

\end{document}